\documentclass[twocolumn,prl]{revtex4}
\begin{document}
\title{Classical and quantum gravity of brane black holes}
\author{Ruth Gregory}
\email{R.A.Gregory@durham.ac.uk}
\author{Simon F. Ross}
\email{S.F.Ross@durham.ac.uk}
\author{Robin Zegers}
\email{Robin.Zegers@durham.ac.uk}
\affiliation{Centre for Particle Theory, Department of  
Mathematical Sciences,
Durham University, South Road, Durham DH1 3LE, U.K.}
\date{\today}
\begin{abstract}
We test the holographic conjecture for brane black holes: that a
full classical 5D solution will correspond to a quantum corrected 4D
black hole. We show that a Schwarzschild-AdS black string in the
bulk can be consistently interpreted as a quantum-corrected black
hole on the brane, but the form of the quantum corrections is unlike
what we would expect. The stress tensor extracted from the bulk
solution does not have a thermal component corresponding to Hawking
radiation outside the black hole. We compare this strong coupling
prediction to a weak coupling calculation to study the differences
in detail. We comment on implications for asymptotically flat
black holes and for black holes localised in the extra dimension.
\end{abstract}
\pacs{PACS numbers: }
\maketitle

In exploring consequences of any quantum theory of gravity
it is the nonperturbative questions that give the most
fascinating opportunities for unexpected physical consequences.
Black holes in particular have provided an extremely fruitful
background for testing our understanding of quantum effects 
in gravity. While it has been known for some time that black
holes emit Hawking radiation \cite{Hawking}, the consequences
of that radiation remain unproven. String
theory has made huge advances in our understanding of
black hole thermodynamics, but as yet is unable to access
the highly non-supersymmetric Schwarzschild black hole.
Clearly, any progress in understanding this physically
relevant case would be extremely important.

Braneworlds are a framework in which the
existence of large extra dimensions is allowed via a 
mechanism which confines standard model physics to a 
slice in spacetime, thus introducing potential hierarchies
in interactions, as well as modifications of gravity
at small (and sometimes large) scales. The brane 
typically warps the bulk spacetime, and in the
Randall-Sundrum (RS) model \cite{RS}, is a slice through
five dimensional anti-de Sitter spacetime. The RS model
makes specific predictions for cosmology
and the LHC dependent on the 5D AdS curvature scale.
But the RS model has another interesting implication:
by taking the near horizon limit of a stack of D3-branes,
the RS model can be thought of as cutting off the spacetime
outside the D-branes; the AdS curvature of the RS bulk
is therefore given rather precisely in terms of the 
D3 brane charge and the string scale. 
Thus, from AdS/CFT \cite{malda}, we might expect a 
parallel between classical branworld gravity, and
quantum corrections on the brane.

There have been several attempts to utilize this relation, in the
context of cosmology \cite{Gubser} and linearized gravity \cite{DL},
for which the evidence is concrete and robust, and in the case of
brane black holes \cite{Tan,EFK}, for which the evidence is more
circumspect, and open to criticism \cite{FRW}.  Briefly, a
cosmological brane is a slice of a bulk black hole spacetime with the
bulk black hole giving rise to a radiation source in the brane
cosmology \cite{BCOS}.  Comparing the temperature of this radiation to
that of a field theory at finite Hawking temperature shows that
these agree up to a factor \cite{Gubser}.  
For linearized gravity, the classical corrections to
the RS brane can be computed from the Lichnerowicz operator, and there
are specific $1/r^3$ corrections to the Newtonian potential \cite{RS}.
These agree {\it precisely} with the 1-loop corrections to the
graviton propagator in quantum gravity \cite{DL}.  Given these
results, it is tempting to suppose that a classical braneworld black
hole solution will correspond to a quantum corrected black hole,
however for this we need an actual solution!

The first attempt to find a braneworld black hole replaced the
Minkowski metric in the RS model with a Schwarzschild metric, giving
rise to an AdS black string \cite{CHR}. This string however suffers
from a classical instability \cite{RG}, so it is not the correct bulk
solution to describe a brane black hole. This instability might
correspond to the thermodynamic instability of the Schwarzschild black
hole via Hawking radiation, although the timescales and nature of the
two instabilities seem to be rather different (see \cite{Fabbri}).

This dual picture led to the conjecture that any nonsingular
braneworld black hole solution must be time dependent
\cite{Tan,EFK,EGK}.  However, the original argument for this relied on
weak coupling calculations, whereas the bulk black hole solution
corresponds to a strongly coupled field theory on the brane, so its
behaviour may be very different. It was argued in~\cite{FRW} that the
quantum-corrected dual description might be consistent with the
existence of static localised black hole solutions. 
It is difficult to construct such solutions explicitly as the system
of equations has too much freedom to be completely classified
analytically \cite{CG}, and the system is very numerically
sensitive. So far, it has been possible to construct static
nonsingular black hole solutions numerically, although these are for
small masses $\lesssim O(\ell^{-1})$ \cite{Kudoh}.

Here we support the point of view of~\cite{FRW} by looking at a
slightly modified RS brane - detuning the brane tension to
subcritical, giving an anti-de Sitter, or Karch Randall (KR) \cite{KR}
braneworld. We consider two KR branes which cross the AdS$_5$ bulk,
both of positive tension, which intersect only formally on the AdS
boundary. There are two types of bulk solution which correspond to a
localised black hole from the braneworld point of view: a black string
stretching between the two branes, or a bulk black hole which is
localised near one brane and does not extend across the whole of the
extra dimension. We focus on the black string, for which an explicit
solution is known which is stable for a range of mass parameters.  In
the regime where it is stable, we would expect this black string to be
the correct solution describing a brane black hole, and even when it
is unstable, the bulk solution is regular, so it should have a
boundary CFT description. We explore the description of this black
string as a quantum-corrected black hole in the brane, and find that a
consistent interpretation exists, but it involves surprising
behaviour.

We start by writing 5D AdS in a general form as a 
foliation over a $4$-dimensional spacetime:
\begin{equation}
g = \Omega^2(u)[ du^2 + \tilde g],
\end{equation}
where $\tilde g$ is a general 4-dimensional metric.  The RS model
takes $\tilde g$ to be Minkowski spacetime, with $\Omega_{RS} =
\ell/u_{RS}$, and $\ell = \sqrt{-6/\Lambda}$ is the 5D AdS length.
The AdS boundary is at $u_{RS}=0$, and the RS brane is at constant
$u_{RS}$.  For a KR brane, we make a simple change to polar
coordinates ($u_{RS} = r\cos\theta$), and put in two branes at $\theta
= \pm \theta_0$.  The AdS boundary (which is excluded from the
spacetime by the introduction of the branes) now corresponds to
$\theta =\pm \pi/2$, and
\begin{equation}
g = \frac{\ell^2}{\cos^2(\theta)} \left [ d\theta^2 +
\frac{\tilde g}{{\tilde \ell}^2} \right ] \label{rmet}
\end{equation}
where $\tilde g$ is now 
an AdS$_4$ geometry with length scale
${\tilde \ell} = \ell \sec \theta_0$. 
Note, these branes both have identical {\it positive} 
tension, $\frac{6\sin\theta_0}{8 \pi G_5 \ell}$,
and the distance between them is {\it finite}. 

Now let $\tilde g$ be the metric of a
Schwarzschild-AdS$_4$ black hole \cite{CK}:
\begin{equation}
\tilde g = -V(r) dt^2 + \frac{dr^2}{V(r)} + r^2 (d\theta^2 + \sin^2
\theta d\phi^2),
\end{equation}
with
\begin{equation} \label{Vr}
V(r) = 1 + \frac{r^2}{\tilde \ell^2} - \frac{2G_4M}{r}. 
\end{equation}
This has a horizon at $r_+$ satisfying $V(r_+)=0$.
This black string stretches between the two branes, and analogous to
the RS black string, we expect that it will exhibit an instability.
This system was analysed in the absence of branes by
\cite{Hirayama:2001bi}, who found stability for $r_+ \gtrsim O(\tilde
\ell)$. There are some technical issues with divergence of 
transverse eigenfunctions in their analysis, however, we have
checked that their conclusion is correct in the presence of the
branes.  Thus, for large $r_+/ \tilde \ell$, the black string is a
suitable brane plus bulk solution for a KR braneworld black hole.  For
small $r_+/\tilde \ell$ (and in particular, in the limit as the brane
cosmological constant goes to zero), this solution is unstable, and
should decay into a localised solution, which would not be of the
simple warped product form we have considered. (Note, in \cite{Wit}, a
different instability for foliations of AdS in terms of compact
negatively curved spaces was discussed. From the braneworld point of
view, this instability corresponds to a scalar tachyonic mode. This
mode is also present in the solutions we consider here; however, it
satisfies the Breitenlohner-Freedman bound for the brane spacetime, so
in the present context, where we are considering a non-compact
braneworld spacetime, it does not imply an instability so long as we
impose the usual asymptotically AdS$_4$ boundary conditions on the
braneworld spacetime.)

We therefore have (for large mass) a stable classical bulk solution
which from  the point of view of the brane 
is exactly a Schwarzschild-AdS$_4$ black hole. 
We would like to understand how this is reconciled with the
viewpoint of \cite{EFK}, that the bulk geometry describes,
from the dual brane/CFT point of view, a quantum-corrected black
hole. One would have expected in general that the back-reaction of the
quantum stress tensor would change the form of the geometry. There is
a non-zero $\mathcal O(N^2)$ stress tensor for the $\mathcal N=4$ SYM
theory on this background, as can be seen by considering the conformal
anomaly, which is
\begin{equation} 
\left\langle
T^\mu{}_\mu \right\rangle = \frac{N^2-1}{32\pi^2} \left (R_{\mu \nu}
R^{\mu \nu} - \frac{1}{3} R^2 \right ) \, .  
\end{equation} 
On an Einstein
space-time such as Schwarzschild-AdS$_4$, we thus have 
\begin{equation}
\left\langle T^\mu{}_\mu \right\rangle = -\frac{N^2-1}{24\pi^2}
\tilde \Lambda^2 = -\frac{3(N^2-1)}{8\pi^2 \tilde \ell^4}\, .  
\end{equation}
So the stress tensor should produce a back-reaction whose effects
would be visible at the order we are considering. Why do we see simply
a Schwarzschild-AdS$_4$ geometry?

The solution is that, as in the discussion of the pure Schwarzschild
black string in \cite{FRW}, the form of the stress tensor we predict
for the strongly-coupled CFT dual to the bulk geometry is very
special. We can evaluate the full stress tensor for the boundary field
theory by using the bulk spacetime and applying the boundary stress
tensor/holographic renormalization approach of
\cite{holo,deHaro:2000xn}. In this approach, we expand the bulk metric
as
\begin{equation}
g = \frac{dz^2}{z^2}+ \frac{1}{z^2} \, [\tilde g_{(0)} + z^2 \tilde
g_{(2)} + z^4 \tilde g_{(4)} + \ldots] \, ,
\end{equation}
and then the stress tensor can be evaluated as \cite{deHaro:2000xn}
\begin{eqnarray}
\label{hologT}
\left\langle T_{\mu \nu} \right\rangle &=& \frac{\ell^3}{4 \pi
G_5} \left [\tilde g_{(4) \mu \nu}  + \frac{1}{8} \left (
\mbox{tr} (\tilde g_{(2)}^2) - \left (\mbox{tr} \, \tilde g_{(2)}
\right )^2 \right ) \tilde g_{(0) \mu \nu} \right. \nonumber \\ &&
\left. - \frac{1}{2}
\left(\tilde g_{(2)}^2 \right )_{\mu \nu} + \frac{1}{4}\tilde g_{(2)
\mu \nu} \mbox{tr} \, \tilde g_{(2)} \right ] \, 
\end{eqnarray}
(where we have ignored some logarithmic terms in the generic
expression which will not contribute in our case).  In our case, the
metric (\ref{rmet}) can be brought into the appropriate form by
writing 
\begin{equation}
\sec\theta = \frac{4 {\tilde \ell}^2 + z^2}{4{\tilde\ell}z}
\end{equation}
so that
\begin{equation}
g = \frac{\ell^2}{z^2} [ dz^2 + \left( 1 + \frac{z^2}{2 \tilde \ell^2} +
\frac{z^4}{16 \tilde \ell^4} \right) \tilde g].  
\end{equation}
Thus we find 
\begin{equation} \label{stress}
\left\langle T_{\mu \nu} \right\rangle = - \frac{3 \ell^3}{64 \pi
G_5 \tilde \ell^4} \tilde g_{\mu\nu}  = - \frac{3N^2\hbar}{32
\pi^2 \tilde \ell^4} \tilde g_{\mu\nu},
\end{equation}
where we use $\hbar G_5 = \frac{\hbar G_{10}}{\pi^3 \ell^5} 
= \frac{\pi \ell^3}{2 N^2}$ in the last step. 
The key point is that this stress
tensor is proportional to the metric on the boundary; the effects of
the back-reaction will therefore be solely to renormalize the
four-dimensional cosmological constant. This special form for the
stress tensor arises directly from the foliated form of the
five-dimensional metric. This is also consistent with the arguments
of \cite{FRW} for the case $\tilde \Lambda=0$: as
$\tilde \ell \to \infty$, $\left\langle T_{\mu \nu} \right\rangle \to
0$, so this leading $\mathcal O(N^2)$ part of the quantum stress
tensor vanishes in this limit. 

Thus, the bulk solution can be consistently interpreted as a
quantum-corrected metric in the dual boundary theory. However, the
form of the boundary stress tensor obtained by this argument is very
different from what we would expect. Our result is independent of the
black hole temperature, whereas we would have expected a component
corresponding to a thermal plasma of CFT degrees of freedom outside
the black hole. The form of a thermal plasma in the strong coupling
CFT is known from AdS/CFT~\cite{Witten:1998zw}. No such contribution
can be seen in (\ref{stress}).

To see the contrast with the expected behaviour in
detail, is instructive to compare the above holographic calculation to
a weak-coupling calculation of the stress tensor of a quantum field on
Schwarzschild-AdS$_4$. We will consider a conformally coupled scalar
field, where an approximate calculation of the quantum stress tensor
on Einstein spaces by Page \cite{Page:1982fm} can be applied. In
Page's approach, we analytically continue to Euclidean signature and
consider the conformally related optical metric, $g_{opt}= \Omega^{-2}
\tilde g$
with $\Omega= V(r)^{-1/2}$, where $V(r)$ is given in (\ref{Vr}).  In
the Euclidean space, to ensure smoothness at the horizon, $\tau$ is
periodically identified with period $\tau \sim \tau + 1/T$, where the
temperature
$T = V'(r_+)/4\pi = (\tilde \ell^2+3r_+^2)(4\pi \tilde \ell^2)
$.
We also write the mass appearing in (\ref{Vr}) in terms of $r_+$ as
$
G_4M= r_+ \left (1 + r_+^2\tilde \ell^{-2} \right )/2 
$.
Page shows that in a Gaussian approximation to the heat
kernel~\cite{Bekenstein:1981xe}, the stress tensor of the scalar field in this
optical metric can be approximated by
$
\left\langle T^{\mu}{}_\nu \right\rangle_{opt} = \frac{\pi^2}{90} T^4 (
\delta^\mu{}_\nu - 4 \delta^\mu{}_0 \delta^0{}_\nu)
$.
The stress tensor in the physical metric can then be determined using
the properties of the field under a conformal transformation. 

Applying this to the Schwarzschild-AdS$_4$ geometry, we find
\begin{widetext}
\begin{equation}
\label{Tbar}
\left\langle {T}^\mu{}_\nu  \right\rangle = \frac{\pi^2}{90(4\pi
r_+)^4} \frac{1}{r^6}  \left [ T^{(1)}(r) \left (\delta^\mu{}_\nu
-4\delta^\mu{}_0 \delta^0{}_\nu \right )  +
3T^{(2)}(r) \delta^\mu{}_0 \delta^0{}_\nu + T^{(3)}(r)
\delta^\mu{}_1 \delta^1{}_\nu \right ] \, , \nonumber
\end{equation}
where we have set
\begin{eqnarray}
T^{(1)}(r) &=& \left (\frac{r-r_+}{r V(r)} \right )^2
\left [ \left (r^2 + 2r r_++3r_+^2 \right ) 
\left (r^4+4rr_+^3-3r_+^4 \right )  
+ \frac{4r_+^2}{\tilde \ell^2} 
\left (3r^6 + 6r^5r_+ + 9r^4r_+^2 +8 r^3r_+^3 +r^2r_+^4 
- 9r_+^6\right ) \right . \nonumber \\
&& \left . +\frac{2r_+^4}{\tilde \ell^4} 
\left (7r^6 + 38r^5 r_+ + 33r^4r_+^2 + 20r^3r_+^3
-17r^2r_+^4-18rr_+^5-27r_+^6 \right )  \right . \nonumber \\
&& \left .-\frac{12r_+^4}{\tilde \ell^6} 
\left (4r^8 + 8r^7r_+ +3r^6r_+^2 - 6r^5r_+^3 
-3r^4r_+^4+5r^2r_+^6+4rr_+^7+3r_+^8 \right ) \right . \nonumber \\
&& \left . -\frac{3r_+^4}{\tilde \ell^8} 
\left (8r^{10}+16r^9r_++24r^8r_+^2+32r^7r_+^3
+13r^6r_+^4-6r^5r_+^5  -r^4r_+^6+4r^3r_+^7+9r^2r_+^8
+6rr_+^9+3r_+^{10} \right )  \right ] \\
T^{(2)}(r) &=& 8r_+^4 \left (3r_+^2 + \frac{6r_+^4}{\tilde \ell^2}
-\frac{2r_+r^3}{\tilde \ell^2} + \frac{3r_+^6}{\tilde \ell^4}
-\frac{2r_+^3r^3}{\tilde \ell^4} - \frac{4r^6}{\tilde \ell^4} \right )\\
T^{(3)}(r) &=& 24 r_+^5 \left (1+\frac{r_+^2}{\tilde \ell^2} \right )
\left (r_+ + \frac{r_+^3}{\tilde \ell^2} + \frac{2r^3}{\tilde \ell^2} \right ) 
\end{eqnarray}
\end{widetext}
This exhibits the expected thermal behaviour.
A useful check of the analysis is to note that $\left\langle T_{\mu
    \nu}\right\rangle$, as given by (\ref{Tbar}), is regular at the
horizon $r=r_+$, as $\left\langle T^0{}_0(r_+) \right\rangle =
\left\langle T^1{}_1(r_+) \right\rangle$.  We can also see that in the
regime where $r, r_+ \ll \tilde \ell $, we recover Page's
result~\cite{Page:1982fm} for the asymptotically flat Schwarzschild
black hole.

This weak coupling result can also be used to consider the 
behaviour for large black holes, which was recently 
considered in \cite{Hemming:2007yq}. In the regime
where $r_+ \gg \tilde \ell$, let us write $r=z r_+$. In terms of $z$,
(\ref{Tbar}) reads, at leading order,
\begin{eqnarray}
\left\langle \bar{T}^\mu{}_\nu (z) \right\rangle &=& \frac{1}{5760
\pi^2} \frac{1}{\tilde \ell^4 z^6} \left [ \frac{-3F(z)}{(1+z+z^2)^2} 
\left (\delta^\mu{}_\nu -4\delta^\mu{}_0 \delta^0{}_\nu \right ) 
\right. \nonumber \\
&& \left. + 24 \left (3-2z^3-4z^6 \right ) \delta^\mu{}_0
\delta^0{}_\nu \right. \nonumber \\
&& \left. +24 \left (1+2z^3 \right ) \delta^\mu{}_1
\delta^1{}_\nu \right ] + {\mathcal O} \left (
\frac{1}{r_+^2 \tilde \ell^2} \right ) \, , 
\end{eqnarray}
where $F$ is a polynomial of order 10, $F(z) = 8z^{10}+ \cdots + 3$.
Thus, we see that at large $r_+$, the quantum stress tensor does not
become large; the factors of $r_+$ cancel out. This shows directly
that quantum corrections remain under control in this regime, as was
argued by other methods in \cite{Hemming:2007yq}. 
 
To sum up: We see that the bulk solution can formally be interpreted
as a quantum-corrected black hole on the brane, but the stress tensor
involved does not have the expected form: it
has the peculiar feature that the correction (a renormalization of the
cosmological constant) is independent of the black hole mass. We also
computed the weak coupling stress tensor, which has the expected form.

It is the strong coupling effect that is particularly intriguing: why
should the black hole apparently not radiate at all in our solution?
This does not seem to be physically sensible.
This special form at strong coupling is
a direct consequence of the fact that the bulk spacetime is foliated
by conformal copies of the Schwarzschild-AdS black hole. This
`translation invariance' means that the classical KK graviton modes
are not excited in the background solution, and geometrically the only
possibility is renormalization of the cosmological constant.  
One might therefore ask if we are considering the correct
solution. After all, black hole solutions are known
not to be unique in 5D. However, we would expect that for
brane black holes with $r_+ > \tilde \ell$, there is a unique stable
regular black hole geometry, and the solution we have found does
describe such a black hole. We therefore believe we
have chosen a good solution. An interesting route to test this would
be to consider a solution not of Einstein gravity, but of
Einstein-Gauss-Bonnet theory. This would introduce inhomogeneities in
the bulk direction, so that even the black string solution would not
be of the simple foliated form. One could then ask if it still
exhibited such peculiar features in the dual description. The main
objection to this is that higher order 
corrections to the Einstein action occur at ${\cal O}(\alpha')$ in the
string action, and our strong coupling computation is in the $N\to
\infty, \alpha'\to0$ limit, in which such corrections should have no effect.

Our calculation shows again that there are real questions about the
interpretation of classical bulk solutions as quantum corrected 
brane solutions. Sufficient puzzles remain that
this will no doubt continue to be a source of lively debate.

\section{acknowledgements}
We would like to thank Roberto Emparan, Alessandro Fabbri, and Nemanja Kaloper
for useful comments. RZ was supported by an EPSRC fellowship.

\providecommand{\href}[2]{#2}\begingroup\raggedright\endgroup


\begin{thebibliography}{10}

\bibitem{Hawking}
S.~W.~Hawking,
Commun.\ Math.\ Phys.\  {\bf 43}, 199 (1975)

\bibitem{RS}
L.~Randall and R.~Sundrum,
Phys.\ Rev.\ Lett.\  {\bf 83}, 4690 (1999),
hep-th/9906064.

\bibitem{malda}
J.~M.~Maldacena,
Adv.\ Theor.\ Math.\ Phys.\  {\bf 2}, 231 (1998),
hep-th/9711200.

\bibitem{Gubser}
S.~S.~Gubser,
Phys.\ Rev.\ D {\bf 63}, 084017 (2001),
hep-th/9912001.

\bibitem{DL}
M.~J.~Duff and J.~T.~Liu,
Phys.\ Rev.\ Lett.\  {\bf 85}, 2052 (2000),
hep-th/0003237.
M.~J.~Duff, J.~T.~Liu and H.~Sati,
Phys.\ Rev.\  D {\bf 69}, 085012 (2004),
hep-th/0207003.

\bibitem{Tan}
T.~Tanaka, 
Prog. Theor. Phys. Suppl. {\bf 148} (2003) 307--316,
gr-qc/0203082.

\bibitem{EFK}
R.~Emparan, A.~Fabbri, and N.~Kaloper, 
JHEP {\bf 08} (2002) 043,
hep-th/0206155.

\bibitem{FRW}
A.~L. Fitzpatrick, L.~Randall, and T.~Wiseman, 
JHEP {\bf 11} (2006) 033,
hep-th/0608208.

\bibitem{BCOS}
P.~Binetruy, C.~Deffayet and D.~Langlois,
Nucl.\ Phys.\ B {\bf 565}, 269 (2000),
hep-th/9905012.
N.~Kaloper,
Phys.\ Rev.\ D {\bf 60}, 123506 (1999),
hep-th/9905210.
P.~Bowcock, C.~Charmousis and R.~Gregory,
Class.\ Quant.\ Grav.\  {\bf 17}, 4745 (2000),
hep-th/0007177.

\bibitem{CHR}
A.~Chamblin, S.~W. Hawking, and H.~S. Reall, 
Phys.  Rev. {\bf D61} (2000) 065007,
hep-th/9909205.

\bibitem{RG}
R.~Gregory,
Class.\ Quant.\ Grav.\  {\bf 17}, L125 (2000),
hep-th/0004101.

\bibitem{Fabbri}
A.~Fabbri and G.~P.~Procopio,
Class.\ Quant.\ Grav.\  {\bf 24}, 5371 (2007),
0704.3728 [hep-th].

\bibitem{CG}
C.~Charmousis and R.~Gregory,
Class.\ Quant.\ Grav.\  {\bf 21}, 527 (2004),
gr-qc/0306069.

\bibitem{EGK}
R.~Emparan, J.~Garcia-Bellido, and N.~Kaloper, 
JHEP {\bf 01} (2003) 079,
hep-th/0212132.

\bibitem{Kudoh}
H.~Kudoh, T.~Tanaka, and T.~Nakamura, 
Phys. Rev. {\bf D68} (2003) 024035,
gr-qc/0301089.
H.~Kudoh, 
Phys.  Rev. {\bf D69} (2004) 104019,
hep-th/0401229.

\bibitem{KR}
A.~Karch and L.~Randall, 
JHEP {\bf 05} (2001) 008,
hep-th/0011156.

\bibitem{CK}
A.~Chamblin and A.~Karch,
Phys.\ Rev.\  D {\bf 72}, 066011 (2005),
arXiv:hep-th/0412017

\bibitem{Hirayama:2001bi}
T.~Hirayama and G.~Kang, 
Phys. Rev. D {\bf 64} (2001) 064010,
hep-th/0104213.

\bibitem{Wit}
N.~Seiberg and E.~Witten, 
JHEP {\bf 04} (1999) 017,
hep-th/9903224.
E.~Witten and S.-T. Yau, 
Adv. Theor. Math. Phys. {\bf 3} (1999) 1635--1655,
hep-th/9910245.

\bibitem{holo}
V.~Balasubramanian and P.~Kraus, 
Commun. Math. Phys. {\bf 208} (1999) 413--428,
hep-th/9902121.
M.~Henningson and K.~Skenderis, 
JHEP {\bf 07} (1998) 023,
hep-th/9806087.

\bibitem{deHaro:2000xn}
S.~de~Haro, S.~N. Solodukhin, and K.~Skenderis, 
Commun.  Math. Phys. {\bf 217} (2001) 595--622,
hep-th/0002230.

\bibitem{Witten:1998zw}
  E.~Witten,
  Adv.\ Theor.\ Math.\ Phys.\  {\bf 2}, 505 (1998)
  [arXiv:hep-th/9803131].

\bibitem{Page:1982fm}
D.~N. Page, 
Phys. Rev.  D {\bf 25} (1982) 1499.

\bibitem{Bekenstein:1981xe}
J.~D. Bekenstein and L.~Parker, 
Phys. Rev. D {\bf 23} (1981) 2850--2869.

\bibitem{Hemming:2007yq}
S.~Hemming and L.~Thorlacius, 
arXiv:0709.3738[hep-th].

\end{thebibliography}
\end{document}